\documentstyle[twoside,fleqn,espcrc2,epsf]{article}
\newcommand{\be}{\begin{equation}}
\newcommand{\ee}{\end{equation}}
\newcommand{\bea}{\begin{eqnarray}}
\newcommand{\eea}{\end{eqnarray}}

\def\bfnabla{\mbox{\boldmath $\nabla$}}
\def\bfSigma{\mbox{\boldmath $\Sigma$}}
\def\bfsigma{\mbox{\boldmath $\sigma$}}

\def\lQ{\Lambda_{\rm QCD}}
\def\al{\alpha}
\def\als{\alpha_{\rm s}}
\def\siml{{\ \lower-1.2pt\vbox{\hbox{\rlap{$<$}\lower6pt\vbox{\hbox{$\sim$}}}}\ }} 
\def\simp{{\ \lower-1.2pt\vbox{\hbox{\rlap{$>$}\lower6pt\vbox{\hbox{$\sim$}}}}\ }}

\newcommand{\Appendix}[1]%
    {%
     \section{#1}%

      }

\begin{document}
\def\siml{{\ \lower-1.2pt\vbox{\hbox{\rlap{$<$}\lower6pt\vbox{\hbox{$\sim$}}}}\ }} 
\def\bfnabla{\mbox{\boldmath $\nabla$}}
\def\bfSigma{\mbox{\boldmath $\Sigma$}}
\def\bfsigma{\mbox{\boldmath $\sigma$}}
\def\als{\alpha_{\rm s}}
\def\al{\alpha}
\def\lQ{\Lambda_{\rm QCD}}
\def\vs{V^{(0)}_s}
\def\vo{V^{(0)}_o}
\newcommand{\ttbs}{\char'134}
\newcommand{\AmS}{{\protect\the\textfont2 A\kern-.1667em\lower.5ex\hbox{M}\kern-.125emS}}

\title{Perturbative quark bound states in pNRQCD\thanks{Invited talk 
given at the International Euroconference QCD 00, Montpellier, France, July
6-13th , 2000. To be published in Nucl. Phys. Proc. Suppl.}}

\author{Nora Brambilla\address{
Institut f\"ur Theoretische Physik, Universit\"at Heidelberg\\ 
Philosophenweg 16, D-69120 Heidelberg, Germany\\
email:n.brambilla@thphys.uni-heidelberg.de}
\thanks{Alexander von Humboldt fellow}}

\begin{abstract}
In the framework of the QCD effective field theory called potential 
Non--Relativistic QCD, we explore  quark--antiquark bound systems that may be 
dominated by the  perturbative interaction and we discuss the extent of validity 
of such a picture. Some phenomenological implications are outlined.
\end{abstract}

\maketitle

\section{INTRODUCTION}
If we have a chance to understand the QCD bound-state dynamics 
without resorting to the lattice or to models \cite{rev}, this 
 is likely to be in the cases of bound systems whose interaction 
is dominated by the perturbative dynamics.
In such cases non-perturbative corrections exist 
but are small and can be parameterized in terms of local or non-local 
condensates. In the following  I will call 
such systems 'Coulombic' or `quasi-Coulombic'.
These  situations are particularly interesting 
not only because, as I will show, they have phenomenological 
relevance, but especially because they allow us to understand much more about QCD.
Here, I will exclusively address such situations.
\par
Typically, systems that may be 'Coulombic or quasi' are bound states composed 
only by heavy quarks (and gluons). In such cases, at least for the lowest 
states, the characteristic radius $\langle r \rangle$ ($r$ being the 
$q\bar{q}$ distance)
 of the system is quite small and this intuitively 
justifies the  idea that they are dominated by the perturbative dynamics: the system
 is too small to probe the confinement effects, which arise at a scale ${1/\lQ}
 \gg \langle r \rangle$.

Here, I will introduce an effective field theory called potential Non--Relativistic QCD
 (pNRQCD)\cite{prepnrqcd,pnrqcd},  that clarifies 
such picture  and allows us to include systematically perturbative and non-perturbative 
contributions.
Since such effective theory is constructed to 
be completely equivalent to QCD, all  the information that we will obtain 
go to deepen our knowledge of bound systems 
in QCD. For  the non-perturbative situation see 
the talk by A. Vairo at this Conference \cite{antoniom} and \cite{pnrqcd,mn}.

\section{THE SCALES of QUARKONIUM}
As it is apparent from the spectra, heavy quarkonia are non-relativistic systems.
Thus, they may  be described 
in first approximation using a Schr\"odinger equation with a potential 
interaction. This amounts to saying  that the heavy quark bound state is characterized
 by three energy scales, hierarchically ordered by the quark velocity 
$v \ll 1$: the quark mass $m$ (hard scale), 
the momentum $mv$ (soft scale (S)),  and 
the binding energy $mv^2$ (ultrasoft scale (US)). In the Coulombic or quasi-Coulombic
situation it is $v \sim \alpha_{\rm s}$.\par
In perturbative calculations,
these scales typically get mixed in the  Feynman diagrams   
 and originate technical complications (the same happens in QED,
 e.g. for positronium). \par
In QCD  a further conceptual complication arises if we take into account 
the existence of the  non-perturbative scale
$\lQ$, at which the non-perturbative effects  become 
dominant. For heavy quarks only the hard scale $m$ is surely bigger than $\lQ$ and 
can be treated perturbatively.\par
The existence of these different scales makes 
 even a purely  perturbative definition of the static $q \bar{q}$ potential 
 not free from complications. Let us consider 
 the 
energy of the static quark sources,
$
E_s(r) \equiv 
  \displaystyle{\lim_{T\to\infty} {i \over T} \ln  \langle W_\Box \rangle},
$
(being $W_\Box$  the static Wilson loop of size ${\bf r} \times T$,  and 
the symbol $\langle ~~ \rangle$ being the average over the gauge fields),
which is usually considered as a definition of the static potential.
 At three loops $E_s$  shows infrared divergences\cite{pot,potpnrqcd}. 
These singularities may indeed be regulated,  upon resummation of a certain 
class of diagrams, which give rise to a sort of dynamical cut-off provided by 
the difference between the singlet and the octet potential. However, such a dynamical 
scale is of the same order of the kinetic energy 
for quarks of large but finite mass and, therefore, should not 
be included into a proper definition of 
the static potential in the sense of the Schr\"odinger equation.
This is similar 
to what happens for the Lamb shift in 
QED at order $1/m^2$. In QCD this effect  calls, even in the definition of the 
static potential,  for a rigorous treatment of the bound-state scales. \par
To address the multiscale dynamics of the heavy quark bound state, 
the concept of effective field theory  turns out to be  not only helpful  but actually
necessary.
\begin{figure}[t]
%\vskip -0.1truecm
\makebox[1.3truecm]{\phantom b}
\put(15,0){\epsfxsize=4.4truecm \epsffile{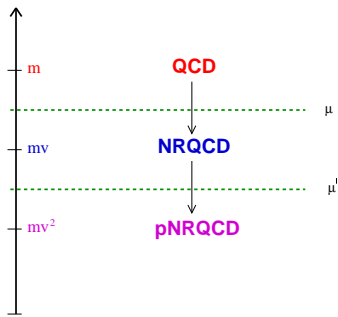}}
%\put(15,210){$V^{(0)}(r)$}
%\put(17,197){\small (GeV)}
\vskip -7.5truemm
\caption{\it  Scales and EFTs  in NR bound state systems.}
\vspace{-6mm}
\label{scales}
\end{figure}
QCD effective field theories 
(EFT) with less and less degrees of freedom, can be obtained 
by systematically integrating out the scales  above the energy we aim to describe. 
This procedure leads ultimately to a {\it field theory 
derived quantum mechanical description} of these systems. The corresponding EFT is called 
pNRQCD \cite{prepnrqcd,pnrqcd}(cf.also\cite{others}). Here, all the dynamical regimes
 are organized in a systematic expansion and the above mentioned  problems are solved. 
In particular we get a procedure to calculate the static potential  beyond two loops and 
to obtain the energy levels beyond next-to-next-to leading order  (${\rm N}^2$LO)\footnote{The 
non-relativistic limit described 
by the Schr\"odinger equation with the static potential is called 'leading order' (LO);
contributions corresponding to corrections of order $v^n$ to this limit are 
called ${\rm N}^n{\rm LO}$. LL means 'leading log'.}.
\vspace{-0.3mm}
\section{NRQCD and pNRQCD}
Since the typical scales of the  system
are widely separated, 
it is possible to perform an expansion of one scale in terms of 
the others and integrate out step by step the scales at higher energy.
 Roughly speaking, first one writes the EFT as a $1/m$ expansion
and integrates out the scale $m$
in QCD: this 
leads to NRQCD. Later, one writes the EFT as an expansion
in the inverse of the soft scale, the so called multipole expansion, and integrates out 
the soft scale in NRQCD: this leads to pNRQCD. See Fig. 1.
The effective theory supplies us with the procedure to make this expansion consistent 
with the ultraviolet behaviour of QCD  and consistent  
with a systematic power counting in the small expansion parameter 
$v$.  Only if the physical system is such that the scales we are integrating out 
are much bigger than $\lQ$, namely $mv \gg \lQ$, then 
it is possible to take advantage of the further simplification 
of performing the calculations in a perturbative expansion in $\alpha_s$ and  to recover a 
perturbative description of the dynamics. This is the  situation we consider here.
 \par
The Lagrangian of NRQCD\cite{NRQCD} can 
be organized in powers of $1/m$. 
In the two-fermions sector it is
 of the type obtained via a  Foldy-Wouthuysen transformation,
but, since we are modifying  the ultraviolet behaviour of the theory, matching coefficients and 
new operators have to be added  in order to mock up the effects of heavy particles  and 
high energy modes  into the low energy EFT.
Since the 
scale of the mass of the heavy quark is perturbative, the scale $\mu$
of the matching from QCD to NRQCD, $mv <\mu < m$, lies also in the perturbative regime.
The integration of  degrees of freedom is done in practice with  a matching procedure
i.e.  by comparing on shell amplitudes,
order by order in ${1/m}$ and in $\alpha_s$,  in QCD and in NRQCD. The difference is 
encoded into  the matching coefficients that typically depend non-analytically on the scale 
$m$ which has been integrated out:
$ c \simeq A \alpha_s (\ln {m\over \mu} +B)$. 
One works typically  in dimensional regularization,
$\overline{MS}$ scheme, and with quark pole masses.\par
After integrating out the soft scale  in NRQCD, pNRQCD  is obtained. The Lagrangian of 
pNRQCD is organized in powers of $1/m$ and ${\bf r}$ (multipole expansion). 
The matching is  done by comparing appropriate off-shell amplitudes in NRQCD and in pNRQCD,
order by order in $1/m$, $\alpha_s$ and order by order in the multipole expansion.
The matching 
coefficients are non-analytic functions of ${\bf r}$  and have 
typically the following structure:
$ V \simeq {\cal V} ({\bf r}, {\bf p}, {\bf S_1}, {\bf S}_2) (A^\prime \ln mr +B^\prime
\ln \mu^\prime r +C)$.
\vskip -0.2truecm 
\section{pNRQCD for $mv \gg \lQ$}

 At the scale of the matching $\mu^\prime$ 
($mv \gg \mu^\prime \gg mv^2, \lQ$) we have still quarks and gluons.
We denote by ${\bf R}\equiv {({\bf x}_1+{\bf x}_2})/2$ the center-of-mass of the $q\bar{q}$ 
system and 
by  ${\bf r\equiv {\bf x}_1 -{\bf x}_2}$ the relative distance. 
The effective degrees of freedom are: $q\bar{q}$ states (that can be decomposed into 
a singlet $S({\bf R},{\bf r},t)$ and an octet $O({\bf R},{\bf r},t)$
under color transformations) with energy of the order of the next relevant 
scale, $O(\Lambda_{QCD},mv^2)$, and momentum\footnote{Although,
for simplicity, we describe the matching between NRQCD and pNRQCD as integrating out 
the soft scale, the relative momentum ${\bf p}$ of the quarks is
still soft.}  ${\bf p}$ of order $O(mv)$,  plus 
ultrasoft gluons $A_\mu({\bf R},t)$ with energy 
and momentum of order  $O(\lQ,mv^2)$. Notice that {\it all the  gluon fields}
 are {\it multipole 
expanded}. The Lagrangian is then  an expansion 
in the small quantities  $ {p/m}$, ${ 1/r  m}$ and in   
$O(\Lambda_{\rm QCD}, m v^2)\times r$.
The pNRQCD Lagrangian is given 
at the leading order in the multipole expansion by\cite{pnrqcd}:
\begin{eqnarray}
& &\hspace{-4mm}
{\cal L}=
{\rm Tr} \Biggl\{ {\rm S}^\dagger \left( i\partial_0 - {{\bf p}^2\over m} 
- V_s(r) -\sum_{n=1} {V_s^{(n)}\over m^n} \right) {\rm S} 
\nonumber \\
& & \!\!\!\!\!\!\!\!\!\!+ {\rm O}^\dagger \left( iD_0 - {{\bf p}^2\over m} 
- V_o(r) - \sum_{n=1} {V_o^{(n)}\over m^n}  \right) {\rm O} \Biggr\}
\nonumber\\
& & \!\!\!\!\!\!\!\!\!\!
 + g V_A ( r) {\rm Tr} \left\{  {\rm O}^\dagger {\bf r} \cdot {\bf E} \,{\rm S}
+ {\rm S}^\dagger {\bf r} \cdot {\bf E} \,{\rm O} \right\} 
\label{pnrqcd0}\\
& &\!\!\!\!\!\!\!\!\!\!\!
   + g {V_B (r) \over 2} {\rm Tr} \left\{  {\rm O}^\dagger {\bf r} \cdot {\bf E} \, {\rm O} 
+ {\rm O}^\dagger {\rm O} {\bf r} \cdot {\bf E}  \right\} -{1\over 4} F^a_{\mu\nu}
F^{\mu \nu a}.  
\nonumber
\end{eqnarray}

All the gauge fields in Eq. (\ref{pnrqcd0}) are evaluated 
in ${\bf R}$ and $t$. In particular ${\bf E} \equiv {\bf E}({\bf R},t)$ and 
$iD_0 {\rm O} \equiv i \partial_0 {\rm O} - g [A_0({\bf R},t),{\rm O}]$. 
The quantities denoted by $V$ are the matching coefficients.
We call $V_s$ and $V_o$ the singlet and octet static matching potentials respectively.
At the leading order in the multipole expansion,
the singlet sector of the Lagrangian gives rise to equations of motion of the 
Schr\"odinger type. The two last lines of (\ref{pnrqcd0})
contain (apart from the Yang-Mills Lagrangian) retardation (or non-potential) effects that 
start at the NLO in the multipole expansion. At this order the non-potential
effects come from the singlet-octet and octet-octet 
interactions mediated by a ultrasoft chromoelectric 
field. 
\begin{center}
\begin{table}
%\vspace{-1.2truecm}
%\makebox[1.3truecm]
\begin{tabular}{|c|c|}
\hline
$ {\rm scale}$ &  bound states \\ \hline
$ \quad m$  \quad  &\qquad \qquad 
  $\psi(2S)$, $\chi_{c0}(1P)$, ... \qquad\qquad \\
          & \qquad \qquad  $\Upsilon(2S)$,   $\chi_{b0}(1P)$, ...
\qquad\qquad\\
          &  long-range hybrids, ... \qquad \qquad \\
$\quad  mv$  \quad & \qquad \qquad
  $\eta_c$, $J/\psi$, $B_c$,... \qquad \qquad\\
          &  short-range hybrids, ...\qquad \qquad \\
$\quad   mv^2$  \quad & \qquad \qquad $t\bar{t}$, $\Upsilon(1S)$,...
\qquad\qquad \\
$ \quad < mv^2$ \quad  & \qquad \qquad QED bound states \qquad \qquad \\
\hline
\end{tabular}
\label{tab1}
%\vskip -1truemm
\caption{\it An indicative  classification of the 
states which can be treated   perturbatively   
up to a scale $ m v^n$  assuming the   criterion 
$m  v^{n}_{\!\!\!\rm pert} / \lQ  \gg  1$.}
\vskip -0.5truecm
\end{table}
\vspace{-0.85truecm}
\end{center}
 Various applications of  pNRQCD have been presented at previous editions
 of this Conference\cite{prepnrqcd,mont}.\par
Recalling that  
 ${\bf r} \sim 1/mv$ and that the operators count like the next relevant 
scale, $O(mv^2,\lQ)$, to the power of the dimension, it follows that  
each term in  the pNRQCD Lagrangian has a definite power counting.  This feature makes 
${\cal L}_{\rm pNRQCD}$ a suitable tool for  bound state calculations: being interested 
in knowing the energy levels up to some power $v^n$, we just need to 
evaluate the contributions
of this size from the Lagrangian. From the power 
counting e.g.,  it follows that the interaction of quarks with ultrasoft 
gluons is suppressed in the Lagrangian
 by $v$ with respect to the LO ( by $g v$ if $mv^2 \gg \lQ$).

In particular, pNRQCD provides us with the way of obtaining the matching 
potentials via 
the matching procedure at any order of the perturbative expansion\cite{pnrqcd}.
In the EFT language the potential is defined upon 
the integration of all the scales {\it up to the ultrasoft 
scale $mv^2$}.  From the matching to NRQCD  in the situation $\lQ \ll mv$
we can easily obtain the  matching potential $V_s$ 
at ${\rm N}^3$LL \cite{potpnrqcd}
$$
\!\!\!\!\!\!\!
V_s(r) = E_s(r)\big\vert_{\rm 2-loop+N^3LL} 
+ C_F {\alpha_{\rm s}\over r} {\alpha^3_{\rm s}\over \pi}
{C_A^3\over 12} \ln {C_A \alpha_{\rm s} \over 2 r \mu^\prime},  
$$
where $E_s$ is the static energy defined in Sec.2 (regulated by the resummation). 
We note that $V_s$ and $E_s$ would coincide 
in QED and that, therefore, this difference here is a genuine QCD feature.
Such difference is switched on  at NLO in the multipole expansion. 
An explicit calculation gives \cite{potpnrqcd}
\begin{eqnarray}
& & \hspace{-7mm} 
V_s(r) \equiv  - C_F {\alpha_{V}(r,\mu^\prime) \over r}, \nonumber\\
& & \hspace{-7mm} 
{\alpha}_{V}(r, \mu)=\alpha_{\rm s}(r)
\left\{1+\left(a_1+ 2 {\gamma_E \beta_0}\right) {\alpha_{\rm s}(r) \over 4\pi}\right. \nonumber\\
& &\hspace{-5mm} 
+{\alpha_{\rm s}^2(r) \over 16\,\pi^2}
\bigg[\gamma_E\left(4 a_1\beta_0+ 2{\beta_1}\right)+\left( {\pi^2 \over 3}+4 \gamma_E^2\right) 
{\beta_0^2}
\nonumber\\
& &\hspace{12mm} 
+ a_2\bigg] \left. + {C_A^3 \over 12}{\alpha_{\rm s}^3(r) \over \pi} \ln{ r \mu^\prime}\right\},
\label{newpot}
\end{eqnarray}
where $\beta_n$ are the coefficients of the beta function ($\alpha_{\rm s}$ is in the $\overline{\rm MS}$ scheme), 
and $a_1$ and $a_2$ were given in \cite{Peter}. 
We see that the interpretation of the potentials as matching coefficients in pNRQCD 
implies that the Coulomb potential is not simply coincident with the static energy $E_s$.
The Coulomb potential turns out to be sensitive to the ultrasoft scale but infrared finite. The same 
happens with the other potentials (like $V_o$ or the potentials that bear corrections in $1/m^n$)
 that can equally be calculated via the matching procedure.
The $\mu^\prime$ scale dependence in the potential is cancelled for any physical 
process by the contribution of the ultrasoft gluons that are cutoff at the scale 
$\mu^\prime$.\par  Then, there are two situations \cite{villefranche}.
 If
$mv\gg mv^2\simp \lQ$, 
the system is described up to order $\alpha_{\rm s}^4$ by a potential which 
is entirely accessible to perturbative QCD. 
Non-potential effects start at order $\alpha_{\rm s}^5\ln \mu^\prime$ \cite{nnnll}. 
We call {\it Coulombic}
 this kind of system. Non-perturbative effects are of non-potential type
and can be encoded into local 
(\`a la Voloshin--Leutwyler\cite{voloshin}) or non-local condensates:
 they are suppressed by powers of  $\lQ/mv^2$ and $\lQ/mv$ respectively. 
If $mv \gg \lQ \gg m v^2$, 
 the scale $mv$ can be still integrated out perturbatively,
giving rise to the Coulomb-type  potential (\ref{newpot}).
Non-perturbative  contributions to the potential arise 
when integrating out the scale $\lQ$ \cite{pnrqcd}.
We call {\it quasi-Coulombic} the  systems  where 
the non-perturbative piece of the potential 
can be considered small with respect to the Coulombic one and treated as a perturbation.
Some levels of $t\bar{t}$, the lowest level of $b \bar{b}$  may be considered Coulombic systems
\footnote{Actually both
 $b\bar{b}$  and $c\bar{c}$
 ground states have been studied in this way\cite{b,yn}},
while the $J/\psi$, the $\eta_c$ and the short-range hybrids may be considered quasi-Coulombic.
The   $B_c$  may be in a boundary situation, see Table 1.
As it is typical in an effective theory,
 only the actual calculation may confirm if the initial
assumption about the physical system was appropriate.

For all these systems it is relevant to obtain a determination of the energy levels 
as accurate as possible in perturbation theory.

\begin{figure}
\makebox[0.8cm]{\phantom b}
\put(-10,40){ S  =}
\put(10,0){\epsfxsize=6truecm \epsfbox{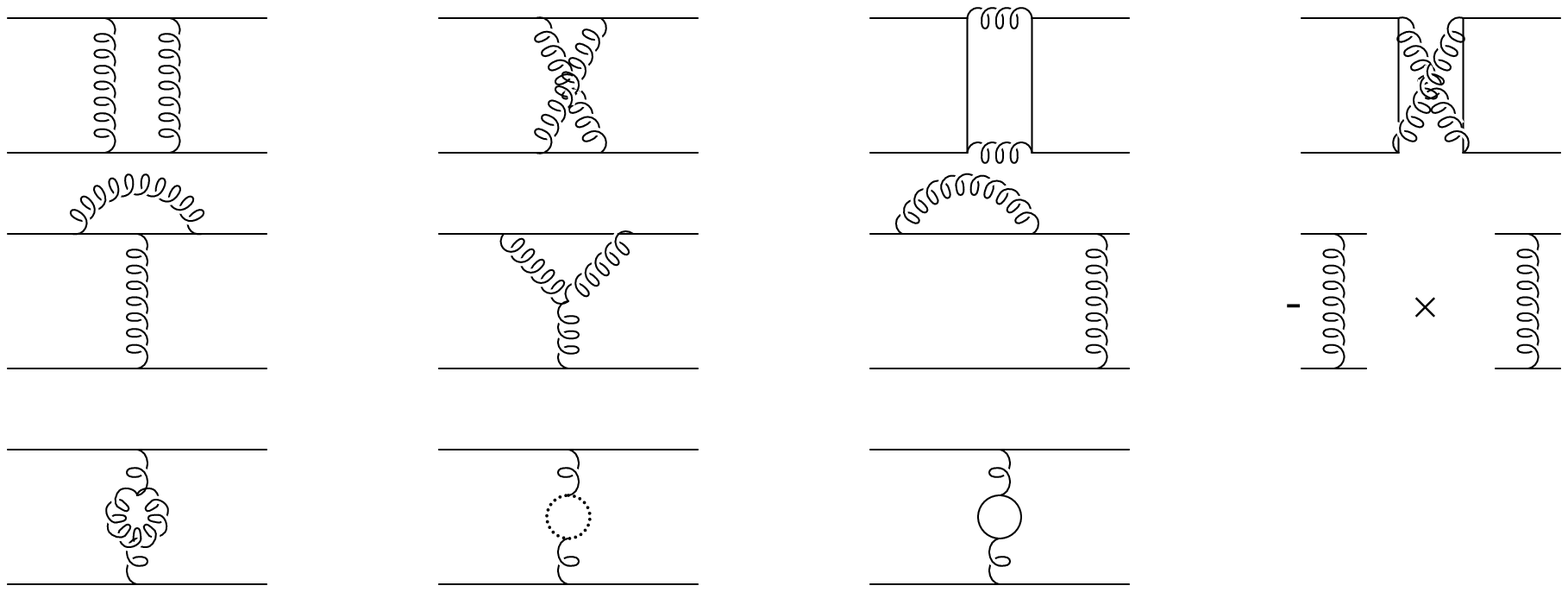}}
\put(-10,-35){ US  = }
\put(65,-36){$\underbrace{\hbox{~~~~~~~~~~~}}$}
\put(50,-50){\small $ 1/(E-V_o-p^2/m)$}
\put(50,-35){\epsfxsize=2.1truecm \epsfbox{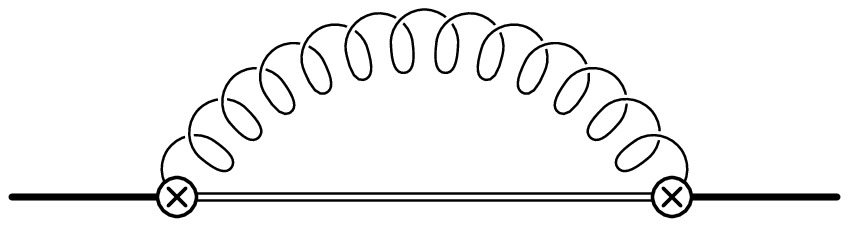}}
\vskip -0.45truecm
\caption{\it Graphs contributing to the spectrum up to ${\rm N}^3$LL.}
\label{graphs}
\vskip -0.6truecm
\end{figure}

\section{CALCULATION of QUARKONIUM ENERGIES at $O(m\alpha_s^5 \ln \alpha_s)$}

The perturbative energy levels of quarkonium are known at $O(m \alpha_s^4)$ from 
\cite{Gupta,b}.
pNRQCD provides us with a well defined way of calculating the energy levels at higher orders,
since the size of each term in the pNRQCD Lagrangian  is
well-defined. 
In order to obtain the leading logs 
at $O(m \als^5)$ in the spectrum, $V_s$ has to be computed at $O(\als^4\ln)$,
$V_s^{(1)}$ at $O(\als^3\ln)$, $V_s^{(2)}$ at $O(\als^2\ln)$ and $V_s^{(3)}$ at $O(\als\ln)$.
The matching to pNRQCD at ${\rm N}^3$LL accuracy was calculated in \cite{nnnll} and the potentials
at the requested accuracy were thus obtained. In Fig.(\ref{graphs}) 
the relevant graphs are shown.  In particular from the last graph in pNRQCD
(the singlet turns into an octet double-line  by emitting an ultrasoft gluon) we extract 
the dependence of $V_s^{(n)}$ on $\mu^\prime$.

The total correction to the energy at $O(m\alpha_s^5 \ln\alpha_s)$ is given by the sum of the 
averaged values of the potentials plus the non-potential effects,
\be
\delta E_{n,l,j}= \delta^{\rm pot} E_{n,l,j}(\mu^\prime)+ \delta^{\rm US} E_{n,l}(\mu^\prime)\, ,
\label{energytotal}
\ee
\bea
&&\!\!\!\!\!\!\!\!\!\!\!\!\!\! \delta E_{n,l,j}^{\rm pot}(\mu^\prime) = E_n {\als^3 \over \pi} 
\left\{{C_A \over 3} \left[{C_A^2 \over 2} 
+4C_AC_F {1\over n(2l+1)}\right. \right.\nonumber \\
& & \!\!\!\!\!\!\!\! \left.\left.
+2C_F^2\left({8 \over n(2l+1)} - {1 \over n^2}\right) \right]
\ln{\mu^\prime \over m\als} \right. \!\!\!\!\!\!\!\!\!\!\!
\label{energy1} \\ 
&&\!\!\!\!\!\!\!\!\!\!\!\!\!\!\!\!\!\!
  +{ C_F^2\delta_{l0} \over 3 n}\left(8\left[C_F-{C_A \over 2}\right]
\ln{\mu^\prime\over m\als} + \left[C_F+{17 C_A \over 2}\right]\ln{\als} \right)
\nonumber \\ 
&& \!\!\!\!\!\!\!\!\!\!\!\!\!\!\!\!\!\!\!\! \left. -{7 \over 3}{ C_F^2C_A \delta_{l0}\delta_{s1} \over n} \ln{\als}  
- {(1- \delta_{l0}) \delta_{s1}C_{j,l} \over l(2l+1)(l+1)n}{ C_F^2C_A \over 2} \ln{\als} \right\} ,
\nonumber
\eea
where $E_n= - mC_F^2\als^2/(4n^2)$ and 
\begin{eqnarray*} 
C_{j,l} = \,\left\{
\begin{array}{ll}
\displaystyle{ -{(l+1)(4\,l-1) \over 2\,l-1}}&\ \ , \, j=l-1 \\
\displaystyle{-1}&\ \ ,\, j=l \\
\displaystyle{{l (4\,l+5)\over 2\,l+3}}&\ \ ,\, j=l+1.
\end{array} \right.
\end{eqnarray*}
The $\ln\als$ appearing in Eq. (\ref{energy1}) come from logs of the type 
$\ln{1/m r}$. Therefore they can be deduced once the 
dependence on $\ln m$ is known. 
  The $\mu^\prime$ dependence of Eq. (\ref{energy1}) 
cancels against contributions from US energies. 
 At the next-to-leading order in the multipole expansion the contribution 
from these scales reads
\begin{eqnarray}
& &\!\!\!\!\!\!\!\!\!\!\!\!\!\!\!\! \delta^{\rm US} E_{n,l}(\mu^\prime) = -i{g^2 \over 3 N_c}T_F \times \!\!\!\!\!\!\!
\label{energyNP} \\
& &\!\!\!\!\!\!\!\!\!\!\!\!\!\!\!\!   \int_0^\infty \!\! dt 
\langle n,l |{\bf r} e^{it(E_n-H_o)} {\bf r}| n,l \rangle \langle {\bf E}^a(t) 
\phi(t,0)^{\rm adj}_{ab} {\bf E}^b(0) \rangle(\mu^\prime), 
\nonumber
\end{eqnarray}
where $H_o = {{\bf p}^2/2m} +V_o$. 

Different possibilities appear depending on the relative size of $\lQ$ with
respect to the US scale $m\als^2$. If we consider that $\lQ \sim m\als^2$ the gluonic correlator in 
Eq. (\ref{energyNP}) cannot be computed using perturbation theory. 
We are still able to obtain all the 
$m \alpha_s^5 \ln({m \alpha_s /m})$ and $m \alpha_s^5 \ln ({m \alpha_s/{\mu^\prime}})$
contributions
 where the
 ${\mu^\prime}$ dependence  cancels now against  the US contributions that have to be evaluated 
 non-perturbatively.

If we consider that $m\als^2 \gg \lQ$, Eq. (\ref{energyNP}) can be computed
perturbatively. Being $m\als^2$ the next relevant scale, the effective role of
Eq. (\ref{energyNP}) will be to replace $\mu^\prime$ by $m\als^2$ (up to finite pieces that 
we are neglecting) in Eq. (\ref{energy1}). Then Eq. (\ref{energytotal}) simplifies to 
\bea
&&\!\!\!\!\!\!\!\!\!\!\!\!\!\!\! \delta E_{n,l,j} = E_n {\als^3 \over \pi} \ln{\als} 
\left\{{C_A \over 3} \left[{C_A^2 \over 2} +4C_AC_F \right.\right. 
{1\over n(2l+1)}\nonumber \\
& & \!\!\!\!\!\!\!\!\!\!\!\!\!\! \left.
+2C_F^2\left({8 \over n(2l+1)} 
- {1 \over n^2}\right) \right] 
 +{ 3 C_F^2\delta_{l0} \over n}\left[C_F+{C_A \over 2}\right] \nonumber \\ 
& & \!\!\!\!\!\!\!\!\!\!\! \left.
- {7 \over 3}{ C_F^2C_A \delta_{l0}\delta_{s1} \over n} 
- {(1- \delta_{l0}) \delta_{s1} \over l(2l+1)(l+1)n}\,C_{j,l}{ C_F^2C_A \over 2} \right\} 
\label{energy2}
\eea    
plus  non-perturbative corrections that can be
parameterized by local condensates \cite{nnnll,voloshin,b} and 
are
of order $ \alpha_s^2 ({\lQ\over m \alpha_s})^2 ({\lQ\over m \alpha_s^2 })^{2}$ and higher.
For the $\Upsilon(1S)$ and $t\bar{t}$  non-perturbative contributions are expected not 
to exceed $100\div 150$ MeV and $10 $ MeV respectively.

The calculation  (\ref{energy2}) 
paves the way to the full ${\rm N}^3$LO order analysis of Coulombic systems and 
is relevant at least for $t\bar{t}$ production and $\Upsilon$
 physics. In the first case it is  a step
forward reaching a 100 MeV sensitivity on the top quark mass for the $t\bar{t}$ cross section 
near threshold to be measured at future Linear Colliders \cite{tt}. In the second case they improve 
our knowledge of the $b$ mass \cite{yn,hoang}.  

In particular, from (\ref{energy2}) we can estimate 
the ${\rm N}^3$LL correction to the energy level of the $\Upsilon(1S)$ and we find 
\begin{eqnarray}
\delta E_{101} & = & {1730\over 81 \pi } 
m_b \alpha_s^4(\mu) \alpha_s(\mu^\prime) \ln{1/\alpha_s(\mu^\prime)}\nonumber \\
& & \simeq (80\div 100)\,  {\rm MeV},
\end{eqnarray}
which appears not to be small.
Corrections from this ${\rm N}^3$LL terms have been calculated for the $t\bar{t}$ cross section 
and $\Upsilon(1S)$ wave function, cf. \cite{pen} and \cite{yn}. 
They also turn out to be sizeable.

Large corrections, however, already show up at NLO and
${\rm N}^2$LO and are responsible for the bad convergence of the perturbative series in terms of the pole mass.
 This will be discussed in the next section.

\section{RENORMALONS, the POLE MASS and the PERTURBATIVE EXPANSION}

The bad convergence of 
the perturbative expansion (for an explicit example  see Sec.7)
can be, at least in part, attributed to  renormalon 
contributions.  The pole mass, thought an infrared safe quantity\cite{kronfeld},  has long 
distance contributions of order $\lQ$\cite{ren}.
Also the static potential is affected by 
 renormalons (see e.g. \cite{ren}). Rephrasing them in the 
effective field theory language of pNRQCD we can say that the Coulomb 
potential suffers from IR 
renormalons ambiguities with the following structure
\begin{equation}
V_s(r) \vert_{\rm IR\, ren} = C_0 + C_2 r^2 + \dots
\label{ren1}
\end{equation}
The constant $C_0 \sim \Lambda_{\rm QCD}$ is known to be cancelled by the IR pole 
mass renormalon ($2 m_{\rm pole}\vert_{\rm IR\, ren} = - C_0$, \cite{ren}). 
Several mass definitions appropriate to explicitly realize this renormalon cancellation 
 have been proposed\cite{mass}. 
Among the others, the $1S$ mass\cite{1s} is defined as half of the  perturbative 
contribution to the $ ^3S_1$ $q\bar{q}$ mass. Unlike the pole mass, the $1S$ mass, containing,
 by construction, half of the total static energy $\langle 2m + V^{Coul}\rangle$, is free of ambiguities of order $\lQ$. Taking e.g. the $\Upsilon(1S)$, 
the $1S$ mass is related to the 
physical $\Upsilon(1S)$ mass by $E(\Upsilon(1S)) =2 m_{1S}+\Lambda_{\Upsilon} $.
$\Lambda_{\Upsilon}$  is the poorly known non-perturbative contribution, which is likely, as
we said, 
to be less than $100\div 150$ MeV.
In the next section we will present an explicit example that shows
 how,  using this mass and thus dealing with  quantities that are infrared safe at order 
$\Lambda_{\rm QCD}$, the pathologies of the perturbative series, 
due to the renormalon ambiguities affecting the pole mass, are cured.\par

It is possible to show 
 that the second infrared renormalon, $C_2 \simeq \lQ^3$, of 
$V_s$ cancels against the appropriate pNRQCD UV renormalon in   
the contribution to the potential originating at NLO in the multipole expansion. 
What remains is the  
 explicit expression for the 
operator which absorbs the $C_2 \sim \Lambda_{\rm QCD}^3$ ambiguity (for  details
 see \cite{pnrqcd}).
An interesting open question is if an 
explicit renormalon subtraction similar to  that one at $O(\lQ)$ can be realized
at the subsequent order $O(\lQ^3)$. This may be relevant since this renormalon is 
related to 
the corrections at ${\rm N}^3$LL discussed in the previous section. Indeed, it is
still an open problem whether the largeness of the ${\rm N}^3$LL corrections
 is an artifact due to our partial knowledge of the contributions at this order,
  or if it is an artifact  
due to the fact that the subsequent renormalon cancellation has to be realized at this order 
or finally if it  is a true signal of the breakdown of the perturbative series.
 To make more definite 
statements one  should know 
the complete ${\rm N}^3$LO or understand the mechanism of cancellation of the second renormalon.

In the next section I will present a concrete example of 
the relevance of the mass renormalon cancellation in order to obtain reliable  phenomenological 
predictions.

\vspace{-0.15truecm}

\section{THE PERTURBATIVE CALCULATION of the $B_c$ MASS}

We consider the perturbative calculation up to order $m\alpha_s^4$ of the  energy of 
the $\bar{b} c$ ground state: this will be relevant  to a QCD determination of the $B_c$ mass
 if this system is Coulombic or at least quasi-Coulombic.
 We also assume 
the $\Upsilon(1S)$ and the $J/\psi$ to be Coulombic or quasi.
The question if these assumptions correspond to the actual systems cannot be settled here.
 On the other hand there is no {\it a priori} reason to rule them out. 
The results will tell us how good were our initial assumptions.

In order to calculate the $B_c$ mass in perturbation theory up to order $\alpha_{\rm s}^4$, 
we only need to consider the following contributions to the potential: the perturbative 
static potential at two loops, the $1/m$ relativistic corrections at one loop,   the spin-independent 
$1/m^2$ relativistic corrections at tree level and the $1/m^3$ correction to the kinetic energy. 
We do not consider  
 $\alpha^5_s \ln{\alpha_s}$ corrections since the mechanism responsible  for this large 
contributions has not yet  been understood. 
Then, we have\cite{bc,b} 
\begin{eqnarray}
& &\!\!\!\!\!\!\!\!\!\!\!
E(B_c)_{\rm  pert} \!=\! m_b \!+\! m_c \! + E_0(\bar{\mu})
\left\{\! 1 \! - {\alpha_{\rm s}(\bar{\mu})\over \pi}
 \right. \!\!\!\!\!\!\!\!\!\!\!\!\! \label{EBc2}\\
& & \!\!\!\!\!\!\!\!\!\!\!\!\!\!\!
 \left[ \beta_0  
l   + {4\over 3} C_A - {11 \over 6} \beta_0 \right] 
+ \left({\alpha_{\rm s}\over \pi}\right)^2 \left[ {3\over 4} 
\beta_0^2 \right.  
l^2   + (2 C_A \beta_0        \nonumber \\
& & \!\!\!\!\!\!\!\!\!\!\!\!\!
 - {9\over 4} \beta_0^2 - {\beta_1\over 4}) 
 l 
- \pi^2 C_F^2  \left( {1\over m_b^2} + {1\over m_c^2} - {6\over m_b m_c} \right) m_{\rm red}^2
\nonumber \\
& & \!\!\!\!\!\!\!\!\!\!\!\!\!\!
+ {5\over 4} \pi^2 C_F^2  \left( {1\over m_b^3} + {1\over m_c^3}\right) m_{\rm red}^3
 + \pi^2 C_F C_A   + {4\over 9} C_A^2 - \nonumber \\
& & \!\!\!\!\!\!\!\!\!\!\!\!
{17 \over 9} C_A \beta_0
+ \left. \left. \left( {181 \over 144} + {1\over 2} \zeta(3) +{\pi^2\over 24} \right) \beta_0^2 
+ { \beta_1 \over 4} + {c\over 8} \right]\right\} ,
\nonumber 
\end{eqnarray}
being $m_{\rm red}= m_b m_c/(m_b+m_c)$,
$l=\ln({2 C_F \alpha_{\rm s} m_{\rm red}/\bar{\mu}})$,
$E_0(\bar{\mu}) = - m_{\rm red} {(C_F \alpha_{\rm s}(\bar{\mu}))^2/2}$ 
and  $\bar{\mu}$ the scale around 
which we expand $\alpha_s(r)$.

If we use here the pole masses $m_b=5$ GeV, $m_c = 1.8$ GeV and 
$\bar{\mu}=1.6$ GeV,   
then we obtain $E(B_c)_{\rm  pert} \simeq 6149$ MeV $\simeq 6800 - 115  - 183 - 353$ MeV, 
where the second, third and fourth figures are the corrections of order $\alpha_{\rm s}^2$,  
$\alpha_{\rm s}^3$ and $\alpha_{\rm s}^4$ respectively. The series turns out to be very badly
convergent.
This reflects also in a strong dependence on the normalization scale $\bar{\mu}$:
at $\bar{\mu} = 1.2$ GeV we would get $E(B_c)_{\rm  pert} \simeq 5860$ MeV, while at 
$\bar{\mu} = 2.0$ GeV we would get $E(B_c)_{\rm  pert} \simeq 6279$ MeV.
The non-convergence of the perturbative 
series (\ref{EBc2}) signals the fact that large $\beta_0$ contributions (coming 
from the 
static potential renormalon) are not summed up and cancelled against the pole masses.  
In order to obtain a well-behaved 
perturbative expansion, we use, now,
the so-called $1S$ mass.   
\begin{figure}[htb]
\vskip -0.1truecm
\makebox[-1.1truecm]{\phantom b}
\put(70,0){\epsfxsize=6truecm\epsffile{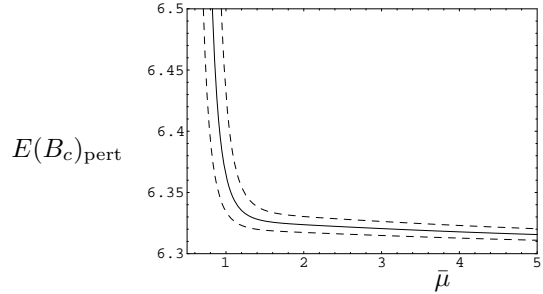}}
\put(40,50){$E(B_c)_{\rm pert}$}
\put(200,1){$\bar{\mu}$}
\vskip -10truemm
\caption{\it $E(B_c)_{\rm pert}$ as a function of $\bar{\mu}$ at $\Lambda_{\overline{\rm
      MS}}^{N_f=3}=300$ MeV (continuous line).  The dashed lines refer 
      to $\Lambda_{\overline{\rm MS}}^{N_f=3}=250,350$ MeV respectively.}
\vspace{-1mm}
\label{plot}
\end{figure}
We consider the perturbative contribution (up to order $\alpha_s^4$) 
of the $\, ^3S_1$ levels of charmonium and bottomonium:
$$
E(J/\psi)_{\rm  pert}= f(m_c); \quad  E(\Upsilon(1S))_{\rm  pert}=f(m_b), 
$$
which are respectively a function of the $c$ and the $b$ pole mass and can be read off from 
Eq. (\ref{EBc2}) in the equal-mass case, adding to it the spin-spin interaction energy: 
$m(C_F\alpha_{\rm s})^4/3$. We invert these relations in order to obtain the pole masses 
as a formal perturbative expansion depending on the $1S$ masses. Finally, we insert the expressions 
$m_c=f^{-1}(E(J/\psi)_{\rm  pert})$ and $m_b=f^{-1}(E(\Upsilon(1S))_{\rm  pert})$ in  Eq. (\ref{EBc2}). 
At this point we have the perturbative mass of the $B_c$ as a function of the  
$J/\psi$ and $\Upsilon(1S)$ perturbative masses.
If we identify the perturbative masses $E(J/\psi)_{\rm  pert}$, $E(\Upsilon(1S))_{\rm  pert}$ with the physical 
ones,  
then the expansion (\ref{EBc2}) depends only on the scale $\bar{\mu}$.
 The perturbative series turns out to be reliable  
for values of $\bar{\mu}$ bigger than $(1.2 \div 1.3)$ GeV and 
lower than $(2.6 \div 2.8)$ GeV. 
For instance, $E(B_c)_{\rm  pert} = 6278.5 +  35 + 6.5 + 5.5$ MeV 
at the scale $\bar{\mu}=1.6$ GeV. 
Therefore, 
we obtain now a better convergence of the perturbative expansion and   
a stable determination of the perturbative mass of the $B_c$.
 This fact seems to support  the $B_c$ {\it being} indeed 
a Coulombic or quasi-Coulombic system.
By varying $\bar{\mu}$ from 1.2 GeV to 2.0 GeV and $\Lambda_{\overline{\rm MS}}^{N_f=3}$ 
from 250 MeV to 350 MeV and by calculating the maximum variation 
of $E(B_c)_{\rm  pert}$ in the given range of parameters, we get 
as our final result  
\begin{equation}
E(B_c)_{\rm  pert} = 6326^{+29}_{-9}\, {\rm MeV}. 
\label{ebc}
\end{equation}
As a consequence of the now obtained good behaviour of the perturbative series 
in the considered range of parameters, the result appears stable with respect 
to variations of $\bar{\mu}$ (see Fig. \ref{plot}) and,
 therefore, reliable from the perturbative point 
of view. It  represents a rather clean prediction 
of the lowest mass of the $B_c$. 
Notice that all the existing predictions are 
based either  on potential models or on lattice evaluation (with still  large errors).

Non-perturbative contributions have not been taken into account so far. 
They affect the identification of the perturbative masses 
$E(B_c)_{\rm  pert}$, $E(\Upsilon(1S))_{\rm  pert}$, 
$E(J/\psi)_{\rm  pert}$, with the corresponding physical ones through Eq.(\ref{EBc2}). 
Let us call these non-perturbative contributions $\Lambda_{B_c}$, 
$\Lambda_{\Upsilon}$ and $\Lambda_{J/\psi}$ respectively. 
As discussed before,
they can be of potential or non-potential nature. 
In the latter
 case they can be encoded into non-local condensates or into local condensates.
  Non-perturbative contributions affect the identification with the 
physical $B_c$ mass roughly by an amount $\simeq -  {\Lambda_{J/\psi}/ 2}$ 
$- {\Lambda_{\Upsilon}/2}$  $+ \Lambda_{B_c}$. 
Assuming $|\Lambda_{J/\psi}| \le 300$ MeV, $|\Lambda_{\Upsilon}| \le 100$ MeV 
and $\Lambda_{\Upsilon} \le \Lambda_{B_c} \le \Lambda_{J/\psi}$, the 
identification of our result (\ref{ebc})
with the physical $B_c$ mass may, in principle, 
be affected by uncertainties, due to the unknown non-perturbative 
contributions, as big as $\pm 200$ MeV. However, the different non-perturbative contributions
$\Lambda$ are correlated, 
so that we expect, indeed, smaller uncertainties. If we assume, for instance, 
$\Lambda_{\Upsilon}$ and $\Lambda_{J/\psi}$ to have the same sign, which seems to be 
quite reasonable, then the above uncertainty reduces to $\pm 100$ MeV. 
 This would confirm, indeed, that 
the effect of the non-perturbative contributions on the result of 
Eq. (\ref{ebc}) is not too large.

\vskip -5truecm

\section{CONCLUSION and OUTLOOK}
We have shown that  for heavy quark bound systems in the situation  $\lQ \siml
 mv^2$,  the energy levels 
turn out to be calculable in perturbation theory  plus local or non-local condensates.
pNRQCD  provides  us with the appropriate tool to calculate  these energy levels
at the  desired order in perturbation theory.
The obtained ${\rm N}^3$LL contributions turn out to be sizeable. A complete 
calculation at ${\rm N}^3$LO is required to settle this issue\cite{sumino}.  
For the practical use of the obtained perturbative series,  renormalon 
cancellation mechanisms  seem to be important. This may be relevant also at 
order $\alpha_s^5\ln{\alpha_s}$. In particular, I have discussed how
a renormalon-free definition 
of the quark masses  improves considerably the behaviour of the perturbative series
on the concrete example of the $B_c$ mass calculation.

\vskip 0.4truecm
{\bf Acknowledgments}\par
I thank the Organizers for invitation and support and the Humboldt foundation  
for support.I thank A. Vairo for reading the manuscript and comments.  
\vskip 0.6truecm
{\bf S. Peris (U.A. Barcelona)}: {\it In your Schr\"odinger equation what definition of 
the quark mass are you supposed to use?}\par
{\bf N. Brambilla}: {\it  The pole mass. Any other definition of the mass  that does not change the power counting 
can be used, cf. Secs. 6 and 7.}

\end{document}